\documentclass[%
 preprint,
 amsmath,amssymb,
 aps,
]{revtex4-1}
\usepackage{braket}
\usepackage{graphicx}
\usepackage{dcolumn}
\usepackage{bm}
\usepackage[hidelinks]{hyperref}

\newtheorem{theorem}{Theorem}
\def\bZ {\mathbb{Z}}

\begin{document}


\title{Normal form for single-qutrit Clifford$+T$ operators and synthesis of single-qutrit gates}

\author{Shiroman Prakash}
\email{sprakash@dei.ac.in}
\author{Akalank Jain}%
\author{Bhakti Kapur}
\author{Shubangi Seth}
\affiliation{%
 Dayalbagh Educational Institute, \\
 Dayalbagh, Agra, India.
}%

\date{March 9, 2018}

\begin{abstract}
 We study single-qutrit gates composed of Clifford and $T$ gates, using the qutrit version of the $T$ gate proposed by Howard and Vala [M. Howard and J. Vala, Physical Review A 86, 022316 (2012)]. We propose a normal form for single-qutrit gates analogous to the Matsumoto-Amano normal form for qubits. We prove that the normal form is optimal with respect to the number of $T$ gates used and that any string of qutrit Clifford+$T$ operators can be put into this normal form in polynomial time. We also prove that this form is unique and provide an algorithm for exact synthesis of any single qutrit Clifford+$T$ operator. 
\end{abstract}

\pacs{Valid PACS appear here}
\maketitle


\section{Introduction}

Fault tolerant quantum computing based on qudits of prime dimension $d\neq2$ \cite{Gottesman1999} is of considerable interest for both theoretical and practical reasons. At a theoretical level, it has been possible to identify contextuality as a necessary feature of quantum mechanics responsible for the power of quantum computers, via the study of magic state distillation of qudits of odd prime dimension \cite{nature}. At a practical level, high thresholds for magic state distillation have been observed for magic state distillation schemes in odd prime dimension  \cite{ACB, CampbellAnwarBrowne, campbell2014enhanced}. See also, e.g.,  \cite{Howard, DPS1, DPS2} for additional results and motivation for studying qudits of odd prime dimension. 

A few important examples of experimental efforts to implement quantum computing using qudits with dimension greater than two include \cite{mair2001entanglement}, which implements  qutrits (and qudits) encoded in the orbital angular momentum states of photons; \cite{PhysRevA.67.062313}, which considers the implementation of a single-qutrit quantum computer via trapped ions; and \cite{PhysRevLett.105.223601} which demonstrates control and tomography of a three-state artificial atom. 

In principle, the computational power of an ideal quantum computer does not depend on the dimensionality of the qudits used as its basic building blocks; however, many important practical aspects -- in particular, the problem of gate synthesis from a finite set of fault-tolerant elementary gates, which we study in this paper -- depend non-trivially on the dimensionality of qudits used as the basic units of information. 

\subsection{A brief review of gate synthesis}

Much work has been done in the area of gate synthesis for qudits, e.g., \cite{muthukrishnan2000multivalued,  2001quant.ph..8062B, bremner2002practical, bullock2005asymptotically, brennan, brennen2006efficient, Di1, Di2}. It has been shown in \cite{PhysRevA.52.3457, 2001quant.ph..8062B, bremner2002practical} that any multi-qudit gate may be implemented exactly by composing single-qudit gates with any two-qudit gate (known as the elementary two-qudit gate) that creates entanglement without ancillas. Some examples of elementary two-qudit gates include the controlled-increment gate proposed by Brennan \textit{et al.} in \cite{brennen2006efficient}, and the generalized controlled-$X$ gate proposed by Di and Wei in \cite{Di1, Di2}. These authors investigated the synthesis of multi-qudit gates using the proposed elementary two-qudit gate, assuming the ability to implement an arbitrary single-qudit gate. 

However, it is generally not possible to implement arbitrary single-qudit gates in a fault-tolerant manner \cite{PhysRevA.57.127}, so any fault-tolerant quantum computer will be restricted to a certain finite set of elementary gates. A finite set of elementary single-qudit gates (supplemented with an elementary two-qudit gate as discussed above) is sufficient for universal quantum computing if it can be used to efficiently approximate an arbitrary single-qudit unitary to any desired accuracy.

In the canonical model for fault tolerant quantum computing, this elementary gate set is the \textit{Clifford group} \cite{Gottesman1999}, which is generated by the two-qudit controlled-SUM gate, the single-qudit Hadamard ($H$) gate and the single-qudit phase ($S$) gate, as we review in detail below in Section \ref{sec:level1}. The single-qudit Clifford group, generated by $H$ and $S$ gates, is a finite group, and thus cannot be used to approximate arbitrary single-qudit gates. To achieve universal quantum computation, one must supplement the Clifford group by at least one additional gate to promote it to an infinite and dense discrete subgroup of $SU(d)$. In the magic state model \cite{MSD} of quantum computation generalized to qudits of odd prime dimension \cite{ACB, CampbellAnwarBrowne, campbell2014enhanced}, a fault-tolerant implementation of the Clifford group is supplemented (via state injection) with the generalized qudit-version of the  $\pi/8$ gate proposed in \cite{HowardVala} -- which we will henceforth refer to the $T$ gate --  to achieve universal quantum computation. 

We are thus naturally motivated to study exact and approximate gate-synthesis of single-qudit unitary gates from elements of the Clifford group supplemented by $T$ gates -- which defines an infinite discrete group generated by $H$, $S$ and $T$ we call the single-qudit Clifford+$T$ group. In the magic state model, the $T$ gate is much more difficult to implement than gates in the Clifford group. Hence, for synthesizing any given single-qudit gate, one would like to use the least number of $T$ gates possible. 

Efficient approximation of arbitrary unitary gates of arbitrary single \textit{qubit} gates from the single-qubit Clifford+$T$ group is a fascinating problem in which great progress has been made, using ideas from algebraic number theory. See \cite{7056491, Selinger:2015:ECA:2685188.2685198, kliuchnikov2013fast, kliuchnikov2013asymptotically, gosset2014algorithm, RossSelinger2014}  and, e.g., \cite{1532, PhysRevLett.112.140504, ross2015algebraic} and references therein for related work in other gate sets. Normal forms \cite{PhysRevLett.109.190501,MatsumotoAmano, GilesSelinger2013, Gosset:2014:AT:2685179.2685180}, particularly the Matsumoto-Amano normal form \cite{MatsumotoAmano}, for single-qubit Clifford+$T$ gates play an important role in these works. 

\subsection{Summary of the paper}

In this paper, we hope to initiate the program of extending these results to qudits of odd prime dimension, by making the first step of proposing a normal form for single qutrit Clifford+$T$ gates, which possesses the same attractive properties as the Matsumoto Amano normal form for qubits. As a byproduct of our investigations, we will also present an efficient exact synthesis algorithm for single qutrit operators in the Clifford+$T$ group. In conjunction with the Solovay-Kitaev algorithm (e.g., \cite{Dawson:2006:SA:2011679.2011685}), our results also enable approximation of arbitrary single-qudit gates from the Clifford+$T$ gate set. However, it is an important open question as to whether more optimal approximation schemes, based on algebraic number theory, exist.

The paper is organized as follows. In Section II and III, we review the concept of a normal form and the Clifford+$T$ group in $p$ dimensions. In Section IV we present our normal form for single-qutrit Clifford+$T$ gates, and in Section V, we prove its existence and T-optimality. In Section VI we prove the uniqueness of our normal form and give an exact synthesis algorithm; we also discuss the problem of approximate synthesis. In Section VII we provide some concrete examples to show how our normal form can be applied in practice. In Section VIII we end with conclusions and discussion.

\section{Normal forms}
Let us briefly review the concept of a normal form. The single-qudit Clifford+$T$ group is an infinite discrete group generated by the operators $H$, $S$ and $T$ (all of which have natural definitions in $p$ dimensions, given in section \ref{sec:level1}). Any element of the single-qudit Clifford+$T$ group can be represented uniquely as an element of $SU(p)$ -- i.e., as a matrix --  or non-uniquely, as a string of $H$, $S$ and $T$'s (such as $HSHTHSHTTHS$). The string can be converted into a matrix, by multiplying the matrices representing $H$, $S$, and $T$ in the order specified. Two strings are equivalent if they correspond to the same element of the Clifford+$T$ group.

The Matsumoto-Amano normal form defines a subset of strings representing single-qubit Clifford+$T$ operators that has the following attractive properties \cite{GilesSelinger2013}:
\begin{enumerate}
\item \textit{Existence} -- There exists an efficient algorithm for converting any string of Clifford+$T$ operators into an equivalent string in Matsumoto-Amano normal form.
\item \textit{Uniqueness} -- The Matsumoto-Amano normal form for any given Clifford+$T$ operator is unique. 
\item \textit{T-optimality} -- The Matsumoto-Amano normal form is the decomposition of a single-qutrit operator into products of Clifford and $T$ gates that contains the minimum number of $T$ gates.
\end{enumerate}

Below we define a normal form for single \textit{qutrit} operators which also has the above properties. Our proof of uniqueness also translates into an exact synthesis algorithm: i.e., given a matrix in $SU(3)$ we can determine whether or not it is in the single-qutrit Clifford+$T$ group, and efficiently construct a string of $H$, $S$ and $T$ operators (using the minimum number of $T$ gates possible) that corresponds to that matrix.

\section{\label{sec:level1} The Clifford+$T$ group in $p$ dimensions}
Before we present our proposed normal form, it is necessary to first briefly review the definition and elementary properties of the Clifford+$T$ group in $p$ dimensions, where $p$ is an odd-prime.

\subsection{The Clifford group in $p$ dimensions}
Pauli operators in $p$ dimensions \cite{Gottesman1999} generalizing $\sigma_x$ and $\sigma_z$ are defined as 
\begin{eqnarray}
X \ket{k} & = & \ket{k+1} \\
Z \ket{k} & = & \omega^k \ket{k}
\end{eqnarray}
Here, and in what follows, letters $j$ and $k$ denote elements of the finite field $\mathbb Z_p$, with multiplication and addition defined modulo $p$. $\omega$ denotes a $p$th root of unity: $\omega^p=1$. Explicitly, for $p=3$, these gates are given by
\begin{equation}
X  =  \begin{pmatrix}
0 & 0 & 1 \\
1 & 0 & 0 \\
0 & 1 & 0 
\end{pmatrix},~
Z  =  \begin{pmatrix}
1 & 0 & 0 \\
0 & \omega & 0 \\
0 & 0 & \omega^2 
\end{pmatrix}.
\end{equation}

The Clifford group in $p>2$ dimensions acting on a single qudit is generated by $S$ and $H$, which are given by:
\begin{eqnarray}
S & = & \sum_{j=0}^{p-1} \omega^{j(j+1) 2^{-1}} \ket{j} \bra{j} \\
H & = & \frac{1}{\sqrt{p}} \sum_{j=0}^{p-1} \sum_{k=0}^{p-1} \omega^{jk} \ket{j}\bra{k}
\end{eqnarray}
where $2^{-1}$ is an element of $\mathbb Z_p$. (Here we are using the definitions given in \cite{Clark2006}, which differ slightly from \cite{Gottesman1999} in the definition of the generator $S$.)  

Let us present the matrices for these gates for qutrits explicitly:
\begin{equation}
H  =  \begin{pmatrix} 1 & 1 & 1 \\ 1 & \omega & \omega^2 \\
1 & \omega^2 & \omega \end{pmatrix}, ~
S  =  \begin{pmatrix} 1 & 0 & 0 \\ 0 & \omega & 0 \\ 0 & 0 & 1 \end{pmatrix}.
\end{equation}

Note that $Z=H^2 S^{-1} H^2 S$ and $X=H^{-1}ZH$, hence $X$ and $Z$ are elements of the Clifford group.

\subsection{$T$ gates in $p$ dimensions}

The analogues of $T$ gates in odd prime dimensions were defined in \cite{HowardVala} and \cite{CampbellAnwarBrowne}. Here we review the results of \cite{HowardVala} for general $p$, though in this paper, we only consider $p=3$. Let us note that, a more general discussion of diagonal gates in the Clifford hierarchy appears in \cite{new1}.

\subsubsection{$p=3$}

For $p=3$ the group of $T$ gates is isomorphic to $\mathbb Z_9 \otimes \mathbb Z_3$ and has two generators. However, the first generator can be obtained from the second using Clifford operations, so it is sufficient to consider a single $T$ gate, which we take to be
 \begin{eqnarray}
 T & = & \begin{pmatrix} \xi & 0 & 0 \\ 0 & 1 & 0 \\ 0 & 0 & \xi^{-1} \end{pmatrix} \label{canonicalT}
 \end{eqnarray}
 Here $\xi=e^{2\pi i /9}$ is a primitive ninth root of unity.
 
$T^3=Z^2$ is an element of the Clifford group, but $T^2$ is not. However, one can check that $T^2 = \xi ZS A T A^{-1}$, where 
\begin{equation}
A=HS^2HS^2H, \label{Adef}
\end{equation} 
so it suffices to implement only one element the $T$ group. In what follows, we will assume that one can implement either $T$ or $T^2$ with $T$-count 1. In other words, we define the $T$-count of a string to be the total number of disjoint $T$ or $T^2$ operators it contains. For example the $T$-count of $HSHTHSHT^2HSHTS$ is 3.

\subsubsection{$p\geq 5$}
For $p>3$ the group of $T$ gates is isomorphic to $\mathbb Z_p^3$ and has three generators: 
\begin{eqnarray}
W & = & \sum_k w_k \ket{k}\bra{k},~ w_k=\omega^{k(1+2k^2-3k) 12^{-1}} \label{t}\\
U & = & \sum_k u_k \ket{k}\bra{k}, ~ u_k = \omega^{k(1+2k^2-3k) 12^{-1}+k} \label{u}\\
V & = & \sum_k v_k \ket{k}\bra{k},  ~ v_k = \omega^{k(1+2k^2+3k)12^{-1}}. \label{v}
\end{eqnarray}
However it is sufficient to consider only one generator as $U$ and $V$ can be obtained from $W$ using Pauli operations and Clifford operations.  This can be taken to be:
\begin{equation}
T_{(p\geq5)} = \sum_k \omega^{k^3 6^{-1}} \ket{k}\bra{k} 
\end{equation}

Note that for $p>3$ one does not need to use any roots of unity other than powers of $\omega$ to define $T$ gates, so the case of $p=3$ is qualitatively different from the case of general $p>3$. 
(The reason the construction in \eqref{t}-\eqref{v} does not apply to $p=2$ and $p=3$ can be traced back to the fact that $6$ does not have in an inverse in $\mathbb Z_2$ and $\mathbb Z_3$.) 

\subsection{The Clifford group and $SL(2,Z_p)$}

Elements of the Clifford group can be related to elements of $SL(2,\bZ_p)$ and translations in discrete phase space \cite{Wootters1987, Gross}, as shown in \cite{Appleby}. (See also \cite{subgroup}.) Let us briefly review this correspondence.

The group $SL(2,\bZ_p)$ is the set of matrices of the form 
\begin{equation} \begin{pmatrix} a & b \\ c & d \end{pmatrix},\end{equation} with all entries in $\bZ_p$ and $ad-bc=1$. $SL(2,\bZ_p)$ has $(p^2-1)p$ elements (which we will denote by overhats), and can be generated from the two matrices $\hat{S}$ and $\hat{H}$ given by:
\begin{equation}
\hat{S}=\begin{pmatrix} 1 & 0 \\ 1 & 1 \end{pmatrix}, ~\hat{H} = \begin{pmatrix} 0 & -1 \\ 1 & 0 \end{pmatrix}.
\end{equation}

Any element of $SL(2,\bZ_p)$ with $b \neq 0$ can be written as $\hat{S}^m \hat{H} \hat{S}^n \hat{H} \hat{S}^q$ with 
\begin{equation}
n= b, ~ q = (a+1)b^{-1}, ~ m = (d+1)b^{-1}, 
\end{equation} 
and any element of $SL(2,\bZ_p)$ with $d \neq 0$ can be written as $\hat{H} \hat{S}^m \hat{H} \hat{S}^n \hat{H} \hat{S}^q$, where
\begin{equation}
n=  d,~m=d^{-1}(1-b), ~ q=(c+1)d^{-1}.
\end{equation}
Together these two forms cover all elements of $SL(2,\bZ_p)$.


The elements of $SL(2,\bZ_p)$, $\hat{S}$ and $\hat{H}$ are closely related to the elements of $U(p)$, $S$ and $H$. As shown in \cite{Appleby}, up to an un-physical overall phase, members of the Clifford group (a subgroup of $U(p)$) can be written as  
\begin{equation}
C=D_{\vec{\chi}}V_{\hat{F}},
\end{equation} 
where $\vec{\chi} = \begin{pmatrix} x \\ z \end{pmatrix}$ is an element of $\mathbb Z_p^2$, $\hat{F}=\begin{pmatrix} a & b \\ c & d \end{pmatrix}$ is an element of $SL(2,\mathbb Z_p)$, 
\begin{equation}
D_{(x,z)} = \omega^{2^{-1} x z} X^xZ^z
\end{equation} 
are Heisenberg-Weyl displacement operators, and the specific form of $V_{\hat{F}}$ was given in \cite{Appleby} to be:
\begin{equation}
V_{\hat{F}}=
\begin{cases}
\frac{1}{\sqrt{p}} \displaystyle \sum_{j,k=0}^{p-1} \omega^{2^{-1} b^{-1}(ak^2-2jk+dj^2)}\ket{j}\bra{k} & b\neq 0 \\
\displaystyle \sum_{k=0}^{p-1}\omega^{2^{-1} ack^2} \ket{ak}\bra{ k} & b= 0 \\
\end{cases}.
\end{equation}
This is a homomorphism, in that 
\begin{equation}
D_{\vec{\chi_1}}V_{\hat{F}_1} D_{\vec{\chi_2}}V_{\hat{F}_2} \sim D_{\vec{\chi}_1+\hat{F}_1\vec{\chi}_2} V_{\hat{F}_1 \hat{F}_2}.\end{equation}
where $\sim$ denotes equality upto an overall phase. 

The generators $S$ and $H$ can be written as $D_{(0,2^{-1})}V_{\hat{S}}$ and $V_{\hat{H}}$ respectively. One can also check that this is an isomorphism \cite{Appleby}. The Clifford group therefore has $p^3 (p^2-1)$ elements (ignoring overall phases).

\section{A normal form for single-qutrit Clifford+$T$ gates}

We now restrict our attention to $p=3$. To present the normal form, we first define the following subsets of the Clifford group $\mathcal H$, $\mathcal H'$ and $\mathcal P$. 

We define $\mathcal P$  to be the subgroup of $\mathcal C$ generated by $S$, $X$ and $V_{-\mathbf 1}$. Note that \begin{equation}
V_{-\mathbf{1}}=\begin{pmatrix}1 & 0 & 0 \\ 0 & 0 & 1 \\ 0 & 1 & 0 \end{pmatrix}.\end{equation}
$\mathcal P$ is equivalent to the group of all operators of the form $D_{p,q}V_{\pm \hat{S}^n}$ where $p,~q,~n \in \mathbb Z_p$. 

Using the fact that the set $\langle \hat{S},-\mathbf{1} \rangle \subseteq SL(2,\mathbb Z_p)$ consists of all matrices of the form:
\begin{equation}
\begin{pmatrix}
\pm 1 & 0 \\ n & \pm 1 
\end{pmatrix},
\end{equation}
one can check that $\mathcal P$ has $2p^3$ elements, and therefore contains $(p^2-1)/2$ left cosets in the Clifford group, which are those elements $V_{\hat{g}}$, where $g$ is one of the left cosets of $\langle \hat{S},-\mathbf{1} \rangle$ in $SL(2,Z_p)$.

\begin{equation}
\begin{pmatrix}
a & b \\ c & d 
\end{pmatrix} \hat{S}^n = 
\begin{pmatrix}
a+nb & b \\ c+nd & d 
\end{pmatrix}
\end{equation}
Using (23) it is easy to see that the left cosets of $\langle \hat{S, -\mathbf{1}} \rangle$ in $SL(2,\mathbb Z_3)$ can be taken to be matrices of the form:
\begin{equation}
\begin{pmatrix}
-1 & 1 \\
c & d 
\end{pmatrix}~\text{ and} 
\begin{pmatrix}
1 & 0 \\
0 & 1 
\end{pmatrix}
\end{equation}
These left cosets can be represented by the following subset of the Clifford group, which we denote as $\mathcal H$:
\begin{equation}
\mathcal H = \{S^m H S H , \mathbf{1} \}
\end{equation}
where $m=0,~1$ or $2$.   

Let us define $\mathcal H ' = \mathcal H \backslash \mathbf{1}$, and $\mathcal C'=\mathcal C \backslash \mathcal P$. Then from the results above, $\mathcal H'$ contains 3 elements, which we denote as $H'_0=H S H$, $H'_1=SHSH$ and $H'_2=S^2HSH$. 

Using these ingredients, we propose the following normal form for single qutrit Clifford+T operators:
\begin{equation}
(\mathcal T|\epsilon) (\mathcal H' \mathcal T)^* \mathcal H \mathcal P. \label{normalform}
\end{equation}
The above expressions uses a hybrid notation of sets and regular expressions that is hopefully self-explanatory -- so, e.g., $(\mathcal H' \mathcal T)$ denotes a string consisting of any element of the set $\mathcal H'$ followed by $\mathcal T$.

Note that $\mathcal C$, $\mathcal H$, $\mathcal P$, $\mathcal T$ and $\mathcal H'$ are finite sets, so we need not be very concerned with the details of how their elements are represented. Each element of $\mathcal P$ could, for instance, be represented as a different character, and not necessarily as a product of $H$'s and $S$'s. For our purposes it is convenient to regard the sets $\mathcal H'$ and $\mathcal T$ above to be composed of the following elementary ``syllables": $\mathcal H'=\{ H_0',~H_1',~H_2' \}$, and $\mathcal T=\{T, T^2\}$. Then the normal form \eqref{normalform} can be written explicitly as
\begin{equation}
(T|T^2|\epsilon)\left((H'_0|H'_1|H'_2)(T|T^2)\right)^* (\epsilon|H'_0|H'_1|H'_2)(\epsilon|V_{-\mathbf{1}})(\epsilon|S|S^2)(\epsilon|X|X^2)(\epsilon|Z|Z^2).
\end{equation}

\section{Existence and T-optimality}
Following the approach of \cite{GilesSelinger2013}, which elegantly rederives basic properties of the Matsumoto-Amano normal form for qubits, we provide a proof of existence and $T$-optimality of the qutrit Matsumoto-Amano normal form. Let us define $\mathcal T=\{T,T^2\}$.

The following facts follow from the discussion above:
\begin{eqnarray}
\mathcal{C} & = & \mathcal{H}\mathcal{P} \\
\mathcal{C'} & = & \mathcal H' \mathcal{P} \label{two} \\
\mathcal{P}\mathcal{H'} & \subseteq &  \mathcal H' \mathcal P  \label{three}
\end{eqnarray}

$S$ clearly commutes with $T$. We also have $XT = \xi TXZS^2$ for qutrits, so 
\begin{eqnarray}
\mathcal{P}\mathcal{T} &= & \mathcal{T}\mathcal{P}. \label{four}
\end{eqnarray} 

Using these facts we now prove that any Clifford+$T$ operator can be written in the normal form \eqref{normalform}. 

Let $M=C_n T^{a_n} C_{n-1} \ldots C_1 T^{a_1}C_0$ be any Clifford+$T$ operator, where $C_i$ are Clifford operators. Apply to $M$ the following the following two-step process:
\begin{itemize}
\item[Step 1:] If any of the $C_i \in \mathcal P$ (other than the leftmost one $C_n$ and the rightmost one $C_0$), then we can use \eqref{four} to replace $T^{a_{i+1}}C_iT^{a_i}C_{i-1}$ by $T^{a_{i}'} C'$ to obtain an equivalent, shorter, expression, without increasing the number of $\mathcal T$ gates required. 
\item[Step 2:] If $a_{i}'=3$, then we replace $T^3$ by $\omega Z^{2}$.
\end{itemize}

Then simplify the Clifford operators occurring in the string (which can be done using, e.g.,  a finite look-up table), and repeat this two-step process until one obtains an expression of the form $$M=C_n \mathcal T C_{n-1} \ldots C_1 \mathcal T C_0$$ where each $C_i \in \mathcal C'$ for $n>i>0$, $C_0 \in \mathcal C$ and $C_n \in \mathcal C$.

Since $\mathcal{C'}  =  \mathcal H' \mathcal P$, (and $\mathcal{C}=\mathcal H \mathcal P$ for the leftmost operator) using relations \eqref{two},\eqref{three} and \eqref{four}, we can push any elements of $\mathcal P$ occurring in any  $C_i$ to the far right of the expression as follows:
\begin{eqnarray}
\mathcal C' \mathcal T \mathcal C' \ldots & = & \mathcal H' \mathcal P \mathcal T \mathcal H' \mathcal P \ldots\\
                                          & = & \mathcal H' \mathcal T \mathcal P  \mathcal H' \mathcal P\ldots \\
                                          & \subseteq & \mathcal H' \mathcal T  \mathcal H' \mathcal P \ldots
\end{eqnarray}
so that each $C_i \in \mathcal H'$, for $i \neq 0$, $i \neq n$. The leftmost operator $C_n$ satisfies $C_n \in \mathcal H$ and the rightmost operator $C_0$  satisfies $C_0 \in \mathcal H \mathcal P$.

We are finally left with an expression of the form
\begin{equation}
M = \mathcal H \mathcal T \mathcal H' \mathcal T \mathcal H' \ldots \mathcal T \mathcal H \mathcal P .
\end{equation}

In the process of converting any Clifford+$T$ operator into the normal form \eqref{normalform} described above, the number of $T$ gates used may have either decreased or stayed the same. Suppose we were given a string representing a single-qutrit Clifford+$T$ operator that uses the minimal number of $T$ gates possible. We could use the above procedure to rewrite it in the normal form \eqref{normalform}, without increasing the number of $T$ gates. Hence we have shown \textit{existence} and \textit{T-optimality} -- for any string representing a Clifford+$T$ operator, there exists an equivalent string written in the normal form \eqref{normalform} which expresses the operator using the minimum number of $T$ gates possible.

Below, we will also show that the decomposition of any single qudit operator into the normal form \eqref{normalform} is unique, hence the normal form of a given Clifford+$T$ operator necessarily uses the minimum number of $T$ gates possible. 

Let us remark that the above existence proof directly translates into an efficient algorithm for rewriting any string of Clifford+$T$ operators in the proposed normal form, though perhaps it could be optimized.





\section{Uniqueness and exact synthesis}

We also need to show that the normal form defined above is unique -- i.e., A given element of the single-qutrit Clifford+$T$ group cannot be expressed as a string in the form \eqref{normalform} in more than one way. We provide a proofs of uniqueness similar in spirit to the proofs of uniqueness for qubits given in \cite{GilesSelinger2013}. This proof of uniqueness translates into an exact synthesis algorithm, similar to \cite{kliuchnikov2013fast} for qubits. (Note, however, that we have not provided an complete algebraic characterization of Clifford+$T$ operators.)

\subsection{Algebraic Preliminaries}
Let us define the following rings:
\begin{itemize}
  \item $\mathbb Z[1/3] = \{\frac{a}{3^n} | ~a \in \mathbb Z,~ n \in \mathbb N  \}$.
  \item $\mathbb Z[\xi] = \{ a_1 +a_2 \xi + a_3 \xi^2 + a_4 \xi^3 + a_5 \xi^4 + a_6 \xi^5|~a_i \in \mathbb Z \}$. In this ring $\xi^3=\omega$ and $\omega^3=1$. We also have $\omega^2=-1-\omega$. Notice that $(1+2\omega)^2=-3$.
  \item $\mathbb Z[\xi,1/3] = \{ a_1 +a_2 \xi + a_3 \xi^2 + a_4 \xi^3 + a_5 \xi^4 + a_6 \xi^5|~a_i \in \mathbb Z[1/3] \}$. 
\end{itemize}

If $a \in \mathbb Z$, let $\bar{a} \equiv a \mod 3$ denote the corresponding element in $\mathbb Z_3$. It is convenient to denote the elements of $\mathbb Z_3$ as $\{ 0, ~+1,~-1\}$.

We define the parity map $P: \mathbb Z[\xi] \rightarrow \mathbb Z_3$ as follows:
\begin{equation}
P(a_1 +a_2 \xi + a_3 \xi^2 + a_4 \xi^3 + a_5 \xi^4 + a_6 \xi^5) \equiv (\bar{a}_1+\bar{a}_2+\bar{a}_3 +\bar{a}_4+\bar{a}_5+\bar{a}_6).
\end{equation}

Let us denote $\chi=1-\xi$. Note that the norm of $\chi$ is 3 in this ring. We remark that the parity map can be thought of as induced by the equivalence relation $a \sim b$ if $(a-b) = c \chi$, where $a,~b,~c \in \mathbb Z[\xi]$, and is a ring homomorphism. 

We define the following \textit{denominator exponent} functions of $y \in \mathbb Z[\xi, 1/3]$: Let $\eta \in \mathbb Z[\xi]$ satisfy $3 \Big| \eta^n$ for some $n$. Then a denominator exponent of $y$ relative to $\eta$ any non-negative integer $k$ such that $\eta^k y \in \mathbb Z[\xi]$. The \textit{least denominator exponent} of $y$ with respect to $\eta$ is the smallest such $k$, and is denoted as $d_\eta(y)$. For example, $d_{\chi}(\frac{1}{3}(\xi - \xi \omega)=3$ and $d_{\chi}(\frac{1}{3})$ is $6$. We will only consider denominator exponents relative to $\chi$ defined above. 

A denominator exponent of a matrix $M$ with entries in $\mathbb Z[\xi, 1/3]$ is defined to be any non-negative value of $k$ such that all the entries of $\chi^k M$ are in $\mathbb Z[\xi]$. The least denominator exponent of $M$ is the smallest such value, which we denote as $d(M)$. 

If $k$ is a denominator exponent of $M$ relative to $\chi$, then we define $P_k(M) \equiv P( \chi^k M)$. 

\subsection{A proof of uniqueness using the $SU(3)$ representation of Clifford+$T$ operators}

We can provide a fairly simple proof of uniqueness using the $U(3)$ representation of Clifford+$T$ operators. Observe that the 3-dimensional matrix representing any Clifford+$T$ operator clearly has entries only in $\mathbb Z[\frac{1}{3},\xi]$.  

Let us first define the $H'$-count $h(M)$ of a Clifford+$T$ operator $M$ as the number of elements of $H'$ it contains (i.e., the number of $H'_0$, $H'_1$ and $H'_2$ syllables appearing in the string.) Let us also define the equivalence relation $\sim_\mathcal{P}$ on elements of the Clifford+T group as follows: $A \sim_\mathcal{P} B$ if $Ag=B$ for some $g \in \mathcal P$. Then, our proof of uniqueness rests on the following theorem:

\begin{theorem} The least denominator exponent $k$ of a $3\times3$ unitary matrix $M$ representing a Clifford+$T$ operator is related to the $H'$-count $h(M)$  via
\begin{equation}
k= \begin{cases} 
h(M)+2 & h(M) \geq 1 \\
0 & h(M)=0.
\end{cases} 
\end{equation}
and
\begin{equation}
P_k(M) \sim_{\mathcal P} 
\begin{pmatrix}
1 & 1 & 1 \\
1 & 1 & 1 \\
1 & 1 & 1 
\end{pmatrix}
\text{ if  $h(M) \geq 1$, and } P_k(M_{3\times3}) \sim_{\mathcal P} 
\begin{pmatrix}
1 & 0 & 0 \\
0 & 1 & 0 \\
0 & 0 & 1 
\end{pmatrix} \text{if  $h(M)=0$.}
\end{equation}
\label{uniqueness3}
\end{theorem}
The proof of this theorem is given in Appendix \ref{proof}.

Using this fact, it is possible to read off the decomposition of an $3\times 3$ matrix representing a Clifford+$T$ operator into a string of syllables in the normal form \eqref{normalform}. Suppose $M$ has denominator exponent $k\geq 3$. (The case $k=0$ can be handled by a finite lookup table.)
Then, if $d({(T^nH'_i)}^{-1} M <k$, (where $n=0,~1$ or $2$) the leftmost syllables of $M$ are $T^n H'_i$. (It follows from the explicit calculations involved in the proof of Theorem 1 that exactly one of the 6 possibilities must be true if the matrix is a Clifford+$T$ operator.) By removing the leftmost syllables from $M$ (by multiplying by ${(T^nH'_i)}^{-1}$) and repeating this procedure recursively, we can uniquely determine the decomposition of a Clifford+$T$ operator into the normal form \eqref{normalform} from its representation as a $3\times 3$ unitary matrix. Hence, our normal form is unique.

The above proof of uniqueness also translates into an exact synthesis algorithm. As we mentioned earlier, any 3-dimensional unitary  matrix representing any Clifford+$T$ operator clearly has entries only in $\mathbb Z[\frac{1}{3},\xi]$. The converse is not true as Theorem \ref{uniqueness3} indicates, and we have not yet discovered a complete algebraic characterization of qutrit Clifford+$T$ operators. Nevertheless, we can still use the above procedure for exact synthesis: if we were to apply the procedure above to a matrix with entries in $\mathbb Z[\frac{1}{3},\xi]$ that is not an element of the Clifford+T group, it would necessarily fail at some step: either it will be impossible to reduce the denominator exponent of the matrix by multiplying by ${H'_i}^{-1}$ or $ (T^n H'_i)^{-1}$, or upon reaching a matrix with denominator exponent $k \leq 3$ it would not match one of the finite possibilities in a look-up table. The total number of steps before it fails (or succeeds) would be of order $k$, where $k$ is the least denominator exponent of the matrix.

\subsection{Comments on approximate synthesis}
Ref. \cite{CampbellAnwarBrowne} argues, using the results of \cite{Nebe2001, Nebe:2006:SCI:1208720}, that the single-qutrit Clifford+T group is dense in the space of single-qutrit unitaries. Therefore, our exact synthesis algorithm, in conjunction with the Solovay-Kitaev algorithm allows for efficient approximation of arbitrary single-qutrit unitaries.

Let us briefly sketch how this works: Suppose we want to approximate the single-qutrit unitary $U$ in $SU(3)$. The Solovay-Kitaev algorithm, which we do not review here as it is covered in many textbooks, would generate an operator $U'$ in the Clifford+T group, given by a string of $H$, $S$ and $T$ operators, that approximates $U$ to some specified accuracy. This string of $H$ $S$ and $T$ operators would, in general, not be given in the normal form we present here, and would likely contain more $T$ gates than necessary. By converting this string into a normal form, using the methods described in the previous section; or alternatively running our exact synthesis algorithm on $U'$, we would obtain the decomposition of $U'$ into $H$, $S$, and $T$ operators which uses the minimum number of $T$ gates. 

However, the Solovay-Kitaev algorithm itself does have limitations, and it is possible that the unitary $U'$ that it outputs is not optimum -- i.e., there might exist another unitary $U''$ that approximates $U$ as well as $U'$, but the minimal $T$-count of $U''$ is less than the minimal $T$-count of $U'$. Obtaining the best approximation of an arbitrary single-qubit unitary for a given $T$-count, is in general a very difficult problem. However, for diagonal single-qubit unitaries, there exist approximation algorithms that perform much better than the Solovay-Kitaev algorithm (at least in practice) \cite{1532, PhysRevLett.112.140504, ross2015algebraic}, that are based on the methods of algebraic number theory. Generalizing these algorithms to the single-qutrit case would require (ii) a normal form for single-qutrit Clifford+T gates that we have provided in this paper, and, (ii) a simple algebraic characterization of the Clifford+T group, which we have not provided here. A simple-algebraic characterization of our Clifford+T group may or may not exist, and we leave this question to future work.

\section{Examples}
In this section, we provide three concrete examples that illustrate how our results may be used in practice.

\subsection{Minimizing the T-count of a string of operators in the Clifford+T group} Suppose one wants to implement the following sequence of single-qutrit operators, which apparently has $T$-count 5: \begin{equation}
    TSHSHTSXTS^2HSHT^2SXTZSHS
\end{equation}
(This string could, for instance, be the output of the Solovay-Kitaev approximation algorithm for some single-qutrit unitary.)
Using, e.g., a finite-lookup table, one can see that this string is of the form
\begin{equation}
    \mathcal T \mathcal H' \mathcal T \mathcal P \mathcal T \mathcal H' \mathcal T \mathcal P \mathcal T \mathcal C. 
\end{equation}
Applying step 1 of the algorithm to both $\mathcal P$ operators we obtain:
\begin{equation}
    \xi^2 T^2 SXZS^4HSHT^3SXZS^2ZSHS
\end{equation}
Now applying step 2, we obtain 
\begin{equation}
    \xi^2 \omega T^2 SXZSHSHZ^2SXZS^2ZSHS
\end{equation}
We see that this is of the form $\mathcal T \mathcal C$, and so is now in normal form with $T$-count 1. The Clifford operator at the end can be simplified using the $SL(2,\mathbb Z_3)$ representation (or a finite lookup table) to be $D_{(2,0)}V_{\hat{S}^2}=X^2Z^2 S^2$ (up to an overall phase).

\subsection{Synthesis of a matrix in the Clifford+$T$ group} Suppose we want to exactly synthesize the following matrix, whose entries are all elements of the ring $\mathbb Z[1/3,\xi]$.
\begin{equation}
M = \left(
\begin{array}{ccc}
 \frac{\xi ^5}{3}-\frac{\xi }{3} & -\frac{\xi
   ^5}{3}-\frac{\xi ^4}{3}-\frac{2 \xi
   }{3}+\frac{1}{3} & -\frac{\xi
   ^4}{3}-\frac{\xi }{3}-\frac{1}{3} \\
 -\frac{\xi ^5}{3}+\frac{\xi ^4}{3}-\frac{\xi
   ^2}{3}+\frac{\xi }{3} & \frac{\xi
   ^5}{3}-\frac{\xi ^3}{3}+\frac{\xi
   ^2}{3}-\frac{1}{3} & -\frac{\xi
   ^5}{3}-\frac{\xi ^4}{3}-\frac{2 \xi
   ^2}{3}-\frac{\xi }{3}-\frac{1}{3} \\
 -\frac{\xi ^3}{3}+\frac{\xi ^2}{3}-\frac{\xi
   }{3}-\frac{2}{3} & \frac{\xi
   ^5}{3}+\frac{\xi ^3}{3}+\frac{\xi ^2}{3} &
   -\frac{\xi ^4}{3}+\frac{\xi ^3}{3}-\frac{\xi
   }{3}+\frac{1}{3} \\
\end{array}
\right).
\end{equation}
We first see that this matrix has denominator exponent 5, as multiplying it by $(1-\xi)^5$ gives a matrix whose entries are all in $\mathbb Z[\xi]$:
\begin{equation}
    \left(
\begin{array}{ccc}
 2 \xi ^5-2 \xi ^3+5 \xi ^2-4 \xi +3 & -4 \xi
   ^5+8 \xi ^4-9 \xi ^3+5 \xi ^2-2 \xi  & 2 \xi
   ^5-2 \xi ^4+3 \xi ^3-2 \xi +3 \\
 -2 \xi ^5+2 \xi ^3-7 \xi ^2+7 \xi -5 & -3 \xi
   ^5+2 \xi ^4-2 \xi ^3-2 & 5 \xi ^5-6 \xi ^4+8
   \xi ^3-4 \xi ^2+\xi +3 \\
 -8 \xi ^5+7 \xi ^4-4 \xi ^3-5 \xi ^2+5 \xi -6
   & 4 \xi ^5-3 \xi ^4+2 \xi ^3+4 \xi ^2-5 \xi
   +5 & 5 \xi ^5-2 \xi ^4+7 \xi ^2-7 \xi +7 \\
\end{array}
\right).
\end{equation}
hence the matrix has $H'$-count 3, by Theorem 1. We find that by multiplying from the left by $(H_2')^{-1}$ reduces the denominator exponent by $1$, so the left most syllable of $M$ is $H_2'$. Removing this left-most syllable, we have $M=H_2'M'$, with $M'$ given by:
\begin{equation}
   M'= \left(
\begin{array}{ccc}
 \frac{\xi ^5}{3}+\frac{\xi ^2}{3}-\frac{\xi
   }{3}-\frac{1}{3} & -\frac{\xi
   ^5}{3}+\frac{\xi ^3}{3}-\frac{\xi
   }{3}+\frac{1}{3} & \frac{\xi
   ^5}{3}+\frac{\xi ^3}{3}+\frac{\xi
   ^2}{3}-\frac{\xi }{3}+\frac{1}{3} \\
 -\frac{\xi ^5}{3}-\frac{\xi }{3}-\frac{1}{3} &
   \frac{\xi ^5}{3}-\frac{\xi ^4}{3}+\frac{\xi
   ^3}{3}+\frac{\xi ^2}{3}+\frac{1}{3} &
   -\frac{\xi ^5}{3}-\frac{\xi ^4}{3}-\frac{\xi
   ^3}{3} \\
 \frac{\xi ^5}{3}-\frac{\xi ^3}{3}+\frac{\xi
   ^2}{3}-\frac{\xi }{3} & -\frac{\xi
   ^3}{3}-\frac{\xi ^2}{3}-\frac{\xi }{3} &
   -\frac{\xi ^5}{3}-\frac{\xi
   ^4}{3}-\frac{1}{3} \\
\end{array}
\right)
\end{equation}
Multiplying $M'$ from the left by $(TH_1')^{-1}$ reduces its least denominator exponent again, so the leftmost syllable of $M'$ is $T H_1'$. Repeating this process until we are left with a matrix of denominator exponent $2$, we find the following decomposition of $M$ into $H$, $S$ and $T$ operators in our normal form:
\begin{equation}
    M = H_2' T H_1' T H_2' T^2 = S^2HSH T SHSH T S^2HSH T^2
\end{equation}
The operator has $T$-count $3$.

\subsection{Attempt to synthesize a matrix outside the Clifford+$T$ group} Suppose we instead wish to synthesize the following matrix, which differs only very slightly from the previous matrix in the example.
\begin{equation}
\tilde{M}=
\left(
\begin{array}{ccc}
 \frac{\xi ^5}{3}+\frac{\xi }{3} & -\frac{\xi
   ^5}{3}-\frac{\xi ^4}{3}-\frac{2 \xi
   }{3}+\frac{1}{3} & -\frac{\xi
   ^4}{3}-\frac{\xi }{3}-\frac{1}{3} \\
 -\frac{\xi ^5}{3}+\frac{\xi ^4}{3}-\frac{\xi
   ^2}{3}+\frac{\xi }{3} & \frac{\xi
   ^5}{3}-\frac{\xi ^3}{3}+\frac{\xi
   ^2}{3}-\frac{1}{3} & -\frac{\xi
   ^5}{3}-\frac{\xi ^4}{3}-\frac{2 \xi
   ^2}{3}-\frac{\xi }{3}-\frac{1}{3} \\
 -\frac{\xi ^3}{3}+\frac{\xi ^2}{3}-\frac{\xi
   }{3}-\frac{2}{3} & \frac{\xi
   ^5}{3}+\frac{\xi ^3}{3}+\frac{\xi ^2}{3} &
   -\frac{\xi ^4}{3}+\frac{\xi ^3}{3}-\frac{\xi
   }{3}+\frac{1}{3} \\
\end{array}
\right)
\end{equation}
This matrix has denominator exponent $6$. Multiplying it from the left with $(T^n H'_m)^{-1}$ for any combination of $n=0,~1,~2$ and $m=0,~1,~2$ only increases its denominator exponent. Hence our exact synthesis algorithm fails, and we conclude that this matrix is not an element of the single qutrit Clifford+T group (despite the fact that its entries are in $\mathbb Z[1/3,\xi]$.) One could also have reached this conclusion by noting that \begin{equation}
    P_6(\tilde{M}) \neq \begin{pmatrix} 1 & 1 & 1 \\ 1 & 1 & 1 \\ 1 & 1 & 1 \end{pmatrix},
\end{equation} which is a necessary (but probably not sufficient) condition for the matrix to be in the single-qutrit Clifford+T group.

\section{Conclusion and discussion}

In this paper, we defined a normal form for single-qutrit Clifford+$T$ operators that shares the useful properties of the Matsumoto Amano normal form for qubits -- namely, existence, uniqueness and $T$-optimality. We also provide an exact synthesis algorithm for single-qutrit Clifford+$T$ operators. Let us briefly comment on some future directions of research.

Our results, in conjuction with the Solovay-Kitaev algorithm can be used for approximate synthesis as well. However, our result is also an important prerequisite for developing  number-theoretic approximate-synthesis algorithms for single-qutrit Clifford+$T$ operators that would be expected to outperform the Solovay-Kitaev algorithm, as they do in the qubit case \cite{1532, PhysRevLett.112.140504, ross2015algebraic}. 

The results also provide us with a better understanding of structure the Clifford+$T$ group for qutrits, which unlike the qubit case is difficult to visualize. 

The question of optimal gate sets for approximation of single-qubit unitaries was studied more systematically in \cite{Goldengates}, and it would be interesting to generalize the analysis for qutrits. 

Finally, it would also be interesting to generalize this normal form to single-qudit Clifford+$T$ operators of arbitrary odd-prime dimension, although it may be challenging to prove uniqueness for general $p$. 

We hope to address these questions in the future. 

\textit{Note Added:} After our paper was submitted for publication, we received the following preprint \cite{2018arXiv180305047G} which also discusses single-qutrit Clifford+T operators. 

\section*{Acknowledgements:} SP would like to acknowledge the support of a DST INSPIRE Faculty award. 
The authors would like to thank Mark Howard for pointing out some useful references about qudits of odd-prime dimension and Peter Sarnak for discussions. We also thank an
We also acknowledge support of a DST-SERB Early Career Research Award (ECR/2017/001023) during the final stages of completion of this work.

\appendix

\section{Proof of Theorem 1}
\label{proof}
In this appendix we present the proof of Theorem \ref{uniqueness3}. 

Notice that any matrix with entries in the ring $\mathbb Z[\xi,\frac{1}{3}]$ with least denominator exponent $k$ can be written in the form 
\begin{equation}
M=\frac{1}{\chi^k}\left(M_{(0)}+M_{(1)} \chi + M_{(2)} \chi^2 +M_{(3)} \chi^3 + \ldots \right),
\end{equation} where the entries of each $M_{(i)}$ are in $\mathbb Z_3$. For instance $M_{(0)}=P_k(M)$, and $M_{(1)}=P_{k-1}\left(M-M_0/\chi^k\right)$, and so on. This series may in general be infinite. Let us refer to the $M_{(i)}$ as the \textit{residues} of $M$. 

Note that the equivalence relation: $\sim_\mathcal{P}$ induced by right-multiplication of elements of $\mathcal P$ is well-defined on residue matrices, and is particularly simple: $M_{(i)} \sim_\mathcal{P} M_{(j)}'$ if $M_{(i)}$ and $M_{(j)}$ differ by a permutation of columns.

The explicit form for the first few residue matrices of the six ``elementary'' syllables $H'_0 T$, $H'_1 T$, $H'_2 T$, $H'_0 T^2$, $H'_1 T^2$, $H'_2 T^2$ used in our normal form are given by:
\begin{equation}
H'_i T^n \sim_P \frac{1}{\chi^3} \left( \begin{pmatrix} 1 & 1 & 1 \\ 1 & 1 & 1 \\ 1 & 1 & 1 \end{pmatrix}  + \chi \begin{pmatrix} 1 &0 & 2 \\ 1 & 0 & 2 \\ 1 &0 &2 \end{pmatrix} + \chi^2 \begin{pmatrix} 0 & 0 & 1 \\ 0 & 0 & 1 \\ 0 & 0 & 1 \end{pmatrix}+\chi^3 A_{i} + \chi^4 B_{i}+ \chi^5 C_{i}\ldots \right) \label{base},
\end{equation}
where, $i=0,~1$ or $2$, and $n=1,$ or $2$, and entries in all matrices are understood to be in $\mathbb Z_3$, 
\begin{equation}
A_0 = \left(
\begin{array}{ccc}
 0 & 1 & 1 \\
 1 & 0 & 1 \\
 0 & 0 & 2 \\
\end{array}
\right), ~ A_1=\left(
\begin{array}{ccc}
 0 & 1 & 1 \\
 0 & 2 & 0 \\
 0 & 0 & 2 \\
\end{array}
\right), A_2 = \left(
\begin{array}{ccc}
 0 & 1 & 1 \\
 2 & 1 & 2 \\
 0 & 0 & 2 \\
\end{array}
\right),
\end{equation}
\begin{equation}
B_0 = \left(
\begin{array}{ccc}
 1 & 1 & 2 \\
 0 & 1 & 2 \\
 1 & 1 & 0 \\
\end{array}
\right), ~ 
B_1 = \left(
\begin{array}{ccc}
 1 & 1 & 2 \\
 1 & 1 & 1 \\
 1 & 1 & 0 \\
\end{array}
\right), ~
B_2 = 
\left(
\begin{array}{ccc}
 1 & 1 & 0 \\
 0 & 1 & 2 \\
 1 & 1 & 2 \\
\end{array}
\right),
\end{equation}
and, 
\begin{equation}
C_0 = \left(
\begin{array}{ccc}
 1 & 2 & 1 \\
 1 & 2 & 1 \\
 1 & 2 & 2 \\
\end{array}
\right), ~ 
C_1 = \left(
\begin{array}{ccc}
 1 & 2 & 1 \\
 1 & 2 & 0 \\
 1 & 2 & 2 \\
\end{array}
\right), ~
C_2 = 
\left(
\begin{array}{ccc}
 1 & 2 & 1 \\
 1 & 2 & 2 \\
 1 & 2 & 2 \\
\end{array}
\right).
\end{equation}

Let $M$ be a matrix with entries in the ring $\mathbb Z[\xi,\frac{1}{3}]$ whose first four residues satisfy the following properties: 
\begin{enumerate}
  \item[P1:] $M_{(0)}=\begin{pmatrix}1 & 1 & 1 \\ 1 & 1 & 1 \\ 1 & 1 & 1 \end{pmatrix},$
  \item[P2:] $M_{(1)} \sim_\mathcal P \begin{pmatrix} 0 &1 & 2 \\ 0 & 1 & 2 \\ 0 &1 &2 \end{pmatrix},$ 
  \item[P3:] $M_{(2)}$ is of the form $\begin{pmatrix}a & b & c \\ a+m & b+m & c+m \\ a-m & b-m & c-m \end{pmatrix},$ with $a$, $b$, $c$ and $m$ in $\mathbb Z_3$,
  \item [P4:] The sum of entries in each column of $M_3$ are equal each other (as elements of $\mathbb Z_3$). In other words, $M_3$ is of the form: $\begin{pmatrix}a & b & c \\ d & e & f \\ m-a-d & m-b-e & m-c-f \end{pmatrix}.$
\end{enumerate}

By a straightforward calculation using the explicit form of the $H'_iT^n$ given above, one can show that if $M$ satisfies the above 4 properties and has denominator exponent $k$, then $M'=H_i T^n M$ also satisfies the above four properties and has denominator exponent $k+1$. Thanks to particular forms of the first two residues of $H'_iT^n$, no assumptions about any of the higher residues of $M$, such as $M_{(4)}$ or $M_{(5)}$, need to be made in this calculation. 

By equation, \eqref{base}, each $H'_iT^n$ also satisfies these above four properties, and therefore by induction, theorem 1 follows, and also any Clifford + $T$ operator of the form $(\mathcal H' \mathcal T)^*(\mathcal H' \mathcal T)$ has residues satisfying properties P1-P4. 

\bibliography{qudit}

\end{document}